\documentclass[aps,prb,twocolumn,groupedaddress]{revtex4-1}
\usepackage{graphicx} 
\usepackage[dvipdfm]{hyperref}
\hypersetup{colorlinks=true,linkcolor=blue,
  implicit=true,breaklinks=true,pagebackref=true,backref=true,
bookmarks=true,bookmarksnumbered=true,hyperfootnotes=true,debug=true,
  naturalnames=false,citecolor=blue,pdfview=FitH,pdfstartview=FitH,hyperindex=true}
\usepackage{amsmath}
\usepackage{amssymb}

\begin{document}

\title{Evidence for filamentary superconductivity nucleated at antiphase
    domain walls in antiferromagnetic CaFe$_2$As$_2$}
\author{H. Xiao$^{1,\star}$,  T. Hu$^{1,4,\dagger}$, A. P. Dioguardi$^{2}$,
N. apRoberts-Warren$^2$, A. C. Shockley$^2$, J. Crocker$^2$, D. M. Nisson$^2$,
Z. Viskadourakis$^{1}$, Xianyang
Tee$^{3}$, I. Radulov$^{1}$, C. C. Almasan$^{4}$,  N. J. Curro$^{2}$,  C. Panagopoulos$^{1,3}$}

\affiliation{$^{1}$Department of Physics, University of Crete and FORTH,
71003, Heraklion, Greece}
\affiliation{$^{2}$Department of Physics, University of California, Davis, CA 95616, USA}
\affiliation{$^{3}$Division of Physics and Applied Physics, Nanyang Technological
University, 637371, Singapore}
\affiliation{$^{4}$Department of Physics, Kent State University, Kent, Ohio, 44242, USA}

\date{\today}
\begin{abstract}
Resistivity, magnetization and microscopic $^{75}$As nuclear magnetic resonance (NMR) measurements  in the antiferromagnetically ordered state of the iron-based superconductor parent material CaFe$_2$As$_2$ exhibit anomalous features that are consistent with the collective freezing of domain walls.  Below $T^*\approx 10$ K, the resistivity exhibits a peak and downturn, the bulk magnetization exhibits a sharp increase, and $^{75}$As NMR measurements reveal the presence of slow fluctuations of the hyperfine field.    These features in both the charge and spin response are strongly field dependent, are fully suppressed by $H^*\approx 15$ T, and suggest the presence of filamentary superconductivity nucleated at the antiphase domain walls in this material.
\end{abstract}


\maketitle

%
%


\subsection{Introduction}
The interplay among competing ground states of correlated electron systems can give rise to a rich spectrum of emergent behavior.  The iron-based superconductors are particularly noteworthy, and have attracted extensive interest since the discovery of La[O$_{1-x}$F$_{x}]$FeAs in 2008. \cite{LaOFFeAsJACS}  Like the high temperature superconducting cuprates, superconductivity (SC) in the iron arsenides emerges from an antiferromagnetic (AFM) parent state upon doping with excess charge carriers, and the superconducting pairing mechanism may be related to the AFM instability of the parent state. \cite{doping122review,EreminChubokovPRL2008,MazinPairingPRL2008,EreminChubokovPRL2008,Fernandes2010} In both cases the electronic degrees of freedom condense into an unusual coexistence of both AFM and SC order parameters for intermediate dopings. \cite{Alldredge2008,NSBa122incommensurate}   The nature of this coexistence is poorly understood. In general, a subdominant order parameter can emerge locally in regions where the dominant order vanishes or is suppressed.\cite{demlerSCandSDW} 
Recent experiments in the iron arsenides suggest that SC and AFM order parameters are indeed spatially modulated on a microscopic scale.\cite{TakigawaSr122pressure,MolerStripesBaFe2As2PRB}

Among the iron based superconductors, the AFe$_2$As$_2$ (A = Ca, Sr and Ba) materials are of particular interest because large single crystals of these oxygen-free compounds can be easily synthesized.\cite{johrendtPRL2008}  These materials undergo a tetragonal to orthorhombic transition followed by an AFM state upon cooling.\cite{Goldman2008PRB}  Both chemical doping and applied pressure suppress the magnetostructural order of the parent compounds and give rise to SC,  and the phase diagrams are similar
to those of other unconventional superconductors.\cite{doping122review,MonthouxPinesReview}  The CaFe$_2$As$_2$ system is noteworthy because SC is induced at only 0.4 GPa (non-hydrostatic), whereas the Sr and Ba materials require 2.8 and 2.5 GPa, respectively.\cite{seunghoCa122PRL,AmesCa122pressure,Cambridge122pressure2009}

In this paper we present evidence for coexisting filamentary SC and AFM in the undoped parent compound CaFe$_2$As$_2$, in which the SC order remains localized within AFM domain walls (DWs).  A similar phenomenon has been observed in heavy fermion materials,\cite{ParkThompsonRh115review} and enhanced superfluid density has been observed at twin boundaries in Ba(Fe$_{1-x}$Co$_x$)$_2$As$_2$.\cite{MolerStripesBaFe2As2PRB} CaFe$_2$As$_2$ is noteworthy, however, because the SC never achieves bulk long-range order, and surprisingly the filamentary SC appears to be related to the presence of low frequency spin fluctuations.  Measurements of the magnetotransport, magnetization, and nuclear magnetic resonance (NMR) reveal anomalies at $T \approx 10$ K that are suppressed with magnetic field in a manner consistent with the suppression of bulk SC in doped samples. Similar anomalies in the AFM state were reported recently to correlate with the superconducting volume fraction in SrFe$_2$As$_2$.\cite{paglione2011} These features  are subtle in the CaFe$_2$As$_2$, but are manifest in both spin and charge probes.  NMR spectra, spin-lattice ($T_1^{-1}$) and spin-spin ($T_2^{-1}$) relaxation measurements indicate slow dynamics and motional narrowing consistent with the freezing of mobile DWs.\cite{adamCaFe2As2,MazinNature2009}  The resistivity and magnetization both change abruptly at the same temperatures and fields and analysis of the magnetotransport indicates the presence of weakly pinned superconducting filaments.  This evidence suggests that the nucleation of  finite superfluid density at DWs in CaFe$_2$As$_2$ drives the freezing of mobile antiphase domains.  We thus find that this nominally pure stochiometric material can spontaneously become electronically inhomogeneous at the nanoscale as a result of coupled lattice strain, antiferromagnetism and superconductivity.\cite{MillisReview}

\subsection{Experimental Details}

High quality  single crystals of CaFe$_2$As$_2$  with typical dimension  $\sim 2\times$2$\times$0.1 mm$^{3}$ were grown in Sn flux by standard methods.\cite{adamCaFe2As2}  Microprobe analysis indicates a Sn concentration $< 600$ ppm in the bulk.  The in-plane resistivity $\rho_{ab}$ was measured using the electrical contact configuration of the flux transformer geometry.\cite{Jiang:1997ee}  Six electrodes were fabricated on each sample by bonding Au wires to the crystal with H20E epoxy paste and the current $I$ was applied in the $ab-$plane.
The NMR spectra and relaxation measurements were acquired on a single crystal at the upper of the two magnetically split central ($I=+\frac{1}{2} \leftrightarrow -\frac{1}{2}$) resonances  of the $^{75}$As ($I = 3/2$) using standard pulse sequences.

\subsection{Results and Discussion}
\begin{figure}
\centering
\includegraphics[width=1.0\linewidth]{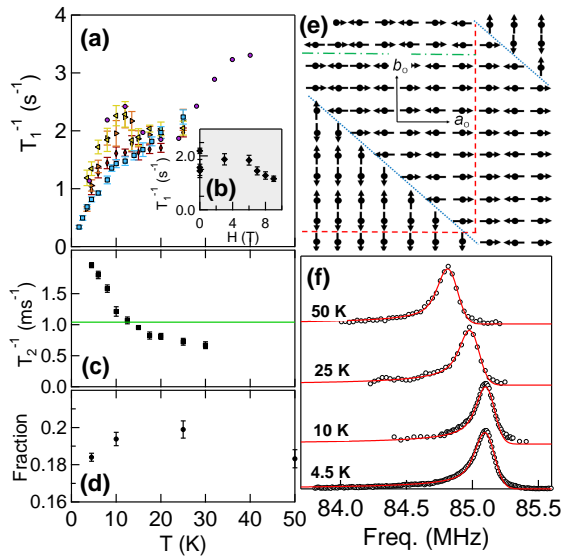}
\caption{\label{fig:NMRsummary} (Color online) (a) $T_1^{-1}$ vs. $T$ for $H_0=0$ ($\bullet$, powder; $\blacklozenge$, satellite), 3 T ($\blacktriangleright$), 6 T ($\blacktriangleleft$) and 9 T ($\blacksquare$) ($H \parallel c$). (b) $T_1^{-1}$ vs. $H_0$ at 8 K ($H \parallel c$).
(c) $T_{2}^{-1}$ for $\mathbf{H}\perp\hat{c}$ at 9 T (solid green line is the like-spin second moment contribution.\cite{abragambook})
(d) Fraction $f = 2\delta/\Delta$ of signal intensity from DWs (see text for details) as a function of $T$.
(e) Schematic of Fe spin orientations in plane indicating different types of domain walls: dotted blue lines are twin boundaries, dashed red are antiphase boundaries along the $\hat{a}$-axis, and dash-dot green lines are antiphase boundaries along the $\hat{b}$-axis.
(f) Spectra of the upper central transition ($\mathbf{H}_{hyp}~ || ~\mathbf{H}~ ||~\hat{c}$) at several temperatures; the red lines are fits as described in the text.}
\end{figure}

$T_1^{-1}$ measurements in the AFM state of CaFe$_2$As$_2$ reveal a small peak at 10 K that was attributed recently to the  presence of slow spin fluctuations possibly associated with DW motion.\cite{adamCaFe2As2}  To test this hypothesis, we measured $T_1^{-1}$  as a function of temperature in different applied field (Fig. \ref{fig:NMRsummary}(a), $H \parallel c$) Surprisingly, we find that the intensity of this peak is suppressed by magnetic fields on the order of 10 T (Fig. \ref{fig:NMRsummary}(b), $H \parallel c$).  We also measured the spin-spin decoherence rate $T_2^{-1}$ versus $T$ for $H\perp c$ (Fig. \ref{fig:NMRsummary}(c)). In this configuration, $T_2^{-2} =\Delta\omega^2_{dip}+(\gamma^2 h_{\rm hyp}^2\tau_c(T))^2$, where $\Delta\omega^2_{dip} = 1.08\times 10^{6}$ sec$^{-2}$ is the temperature independent second moment of the like-spin dipolar couplings among the $^{75}$As for the central transition with $\mathbf{H}||[100]$ (solid line in Fig. \ref{fig:NMRsummary}(c)).\cite{abragambook,PenningtonSlichterT2G}   $T_2^{-1}$ clearly increases sharply below 25 K, revealing the presence of slow fluctuations of the hyperfine field in the $ab$ plane.

In the AFM ordered state, there are two types of  domain boundaries that can give rise to slowly fluctuating hyperfine fields: twin boundaries and antiphase boundaries (see Fig. \ref{fig:NMRsummary}(e)).\cite{MazinNature2009} Twin boundaries emerge to relieve lattice strain and are immobile.\cite{prozorovDW122}  Antiphase DWs have been detected well below $T_N$ in BaFe$_2$As$_2$, and are likely to be mobile depending on the temperature.\cite{PlummerBa122STM}  Both types of DWs give rise to perturbations of the local hyperfine field and could be responsible for the slow dynamics we observe.\cite{DioguardiBa122} The DWs give rise to a low frequency tail in the spectra that is clearly evident in Fig. \ref{fig:NMRsummary}(f). By using the known hyperfine couplings to the As, we model the resonance frequency as a function of position upon crossing such a DW as
 $\omega(x) = \gamma H_0 + \omega_0\tanh(x/\delta)$ where  $x$ is the position perpendicular to the DW, $\delta$ is the DW width,  $\omega_0$ is the frequency arising from the hyperfine field,  $\omega_0 =\gamma H_{\rm hyp} = 19.2$ MHz, and $\gamma H_0 = 65.4$ MHz.\footnote {See Supplemental Material for details of the calculation}
 We fit the spectra and extract the fraction $f = 2\delta/\Lambda$ of the signal intensity arising from DWs, where $\Lambda$ is the domain width.  This portion is roughly independent of temperature down to 10 K, and then it decreases by approximately 5\% (Fig. \ref{fig:NMRsummary}(d)) and appears to be correlated with the slow dynamics observed in $T_1$ and $T_2$. Twin boundaries are most likely the source of the temperature independent contribution.    Motional narrowing from mobile antiphase domain walls may be responsible for the the slow dynamics below 10 K, which may serve to suppress $\delta$ and hence $f$. In this case the correlation time  $\tau_c(T)\gtrsim(\gamma h_{\rm hyp})^{-1}\sim 10^{-6}$ s, where $\gamma$ is the gyromagnetic ratio and $h_{\rm hyp}\sim 1.4$ kOe is the fluctuating component of the hyperfine field in the $ab$ plane at the DW, which agrees with the calculated value at an antiphase boundary.

\begin{figure}
\centering
\includegraphics[trim=0cm 0cm 0cm 0cm, clip=true, width=0.5\textwidth]{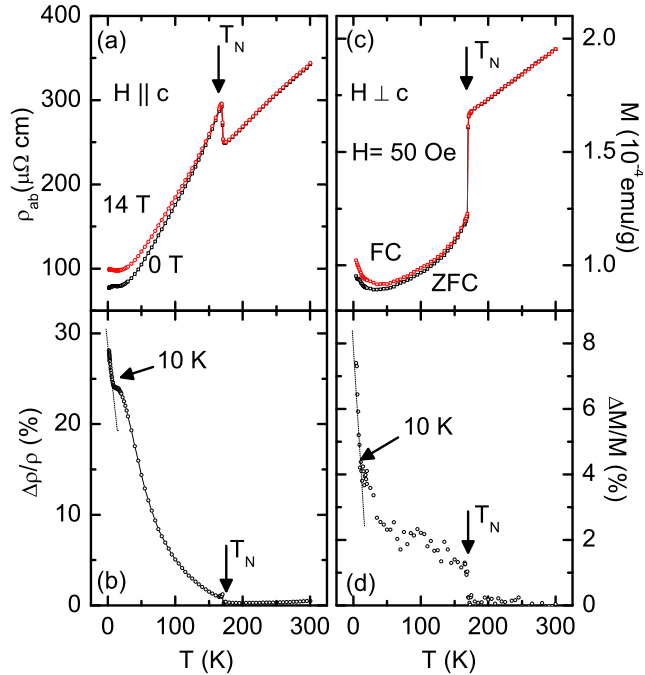}
\caption{ \label{fig:rhoandM}(Color online) (a) Temperature $T$ dependence of the in-plane resistivity $\rho_{ab}$ for applied magnetic fields $H$=0 and
14 T and an applied current $I=$ 1 mA. (b) $T$ dependence of magetoresistivity $\Delta \rho/\rho$ where $\Delta \rho=\rho(14T)-\rho(0)$.
(c) $T$ dependence of ZFC and FC ($H=$ 50 Oe) magnetization $M$ curves. (d) $\Delta M/M$ vs. $T$ where $\Delta M= M(\text{FC})-M(\text{ZFC})$. }
\end{figure}

We now turn to the bulk transport and magnetization experiments. Figure \ref{fig:rhoandM}(a) depicts the temperature dependence of the resistivity $\rho_{ab}$  at $H=0$ and 14 T ($H \parallel c$), which exhibits a discontinuity at $T_N=$169 K consistent with the reported phase transition.\cite{RonningCaFe2As2discovery}
The magnetoresistance $\Delta \rho/\rho=[\rho(14T)-\rho(0)]/\rho(0)$ (Fig. \ref{fig:rhoandM}(b)) is almost zero for $T>T_N$, but increases below $T_N$ and shows a sharp upturn at 10 K.  Figure \ref{fig:rhoandM}(c) shows the zero-field-cooled (ZFC) and field-cooled (FC) magnetization $M$. The hysteresis below $T_N$  suggests the presence of magnetic domains, and the low temperature increase in $M$ may be due to free moments present in the DWs; we estimate the percentage of free moments to be $\sim$7-18.7$\%$ for ordered moments of iron, $\mu_{eff}$=0.3-0.8 $\mu_B$.  Figure \ref{fig:rhoandM}(d) depicts the $T$ dependence of  $\Delta M/M=[M(\text{FC})-M(\text{ZFC})]/M (\text{ZFC})$.  Below 10 K the slope increases dramatically, indicating a collective freezing of the magnetic domains.  It is clear  from these data that the charge transport is strongly coupled to the low temperature anomaly present in the NMR and magnetization data.

\begin{figure}
\centering
\includegraphics[trim=0cm 0cm 0cm 0cm, clip=true, width=0.5\textwidth]{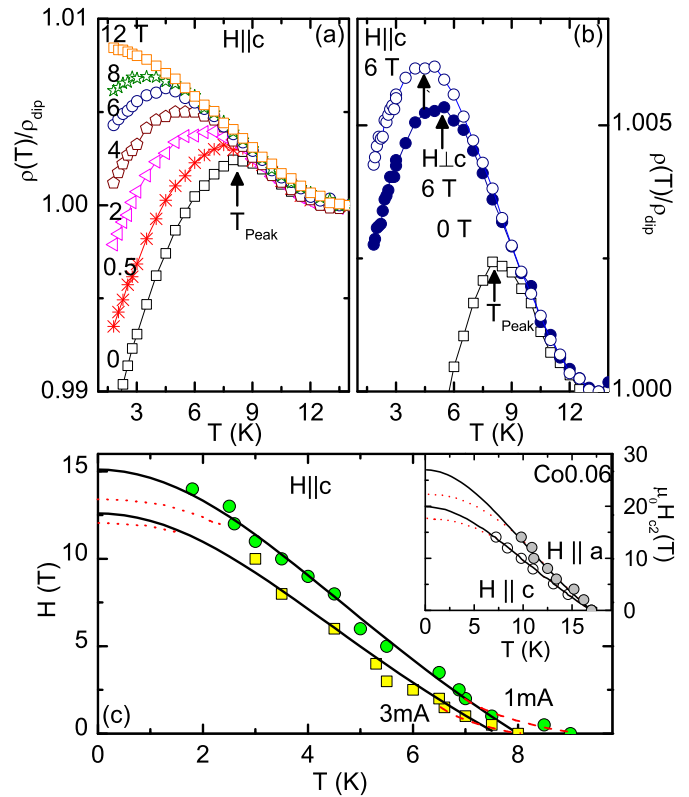}
\caption{ \label{fig:reducedrho} (Color online) (a) $T$ dependence of the reduced resistivity $\rho(T)/\rho_{\text{dip}}$ measured at $H =$ 0, 0.5, 2, 4, 6, 8, 12 T ($H \parallel c$) and $I=$ 3 mA.  (b) $H =$ 0 and 6 T for both $H\parallel c$ and $H \perp c$  with $I=$ 3mA. (c)  $T$ and $H$ dependence of the peak position for $I =$ 1 and 3 mA.
Inset: $\mu_0 H_{c2}-T$ phase diagram of CaFe$_{1.94}$Co$_{0.06}$As$_2$.
(From Ref. \cite{KumarPRB2009}).
The solid black lines are fits to the GL expression for upper critical field; the dotted red lines are fits by WHH relation.}
\end{figure}

In order to highlight the low temperature magnetotransport anomaly,  Fig. \ref{fig:reducedrho}(a) shows the low temperature resistivity normalized by $\rho_{\rm dip}$, the value at the local minimum at $\sim13.5$ K.  Below this temperature the resistivity exhibits a local maximum around 10 K, an observation that is consistent with other reports.\cite{Ni:2008jp, CanfieldPressureCa122} Increasing  the magnetic field reveals a semiconducting  like background with an enhanced resistivity below 10 K.  Below this temperature, however, the resistivity exhibits a field-dependent peak and downturn at lower temperature. The peak position  $T_{\text{peak}}$ (marked with an arrow in Figs. \ref{fig:reducedrho}(a) and \ref{fig:reducedrho}(b)) shifts to lower $T$ with increasing field.  $T_{\rm peak}$  is suppressed with field and  $\rho$ no longer exhibits a peak by $H>12$ T, but continues to rise down to the lowest temperature measured.  The resistivity peak is similar for both $H \parallel c$ and $H \perp c$ (Fig. \ref{fig:reducedrho}(b)) indicating a very small anisotropy consistent with earlier reports in SC samples of this family.\cite{Torikachvili:2009dp}  $T_{\rm peak}$ shifts slightly as a function of the applied field direction, and is smaller for $H \parallel c$.

The field dependence of $T_{\rm peak}$ is summarized in Fig. \ref{fig:reducedrho}(c) and bears a striking resemblance to the reported upper critical field $H_{c2}$ vs $T$ phase diagram for SC CaFe$_{1.94}$Co$_{0.06}$As$_2$ (upper inset).\cite{KumarPRB2009}   We find $H$ is nearly linear in $T_{\rm peak}$ with a slope of -2.4 T/K for both current values.  Linear extrapolations to zero temperature yields critical field values of 18.3 and 16.7 T, respectively, representing the applied fields necessary to completely suppress the downturn in $\rho$. In fact, our observations are consistent with the development of filamentary superconductivity, in which only a small volume fraction of the material becomes superconducting.  This fraction is responsible for the suppression of resistivity, but is not sufficiently extended spatially to lead to a diamagnetic signature or a completely zero resistance state.  The facts that $T_{\rm peak}$ is on the same scale as $T_c$ in fully superconducting samples and that both temperatures are suppressed with field in similar manners suggest that the phenomena we observe are not associated with an impurity phase.  This observation is  supported by fits
to the Ginzburg-Landau (GL) expression for the upper critical field, $H_{c2}(T)=H_{c2} (0) [1-(T/T_c)^2]/[1+(T/T_c)^2]$ (solid black lines, yielding $H_{c2}(0)=15.1$ T, $T_{c}(0)=8$ K for 1 mA and 12.6 T and 7.6 K for 3 mA) and the  Werthamer-Helfand-Hohenberg (WHH) relation, $H_{c2}(0)=-0.7T_c(dH_{c2}/dT_c)$ (dotted red lines, yielding 13.3 T and 12 T for 1 mA and 3 mA, respectively). The value of $H_{c2}(0)$ obtained from GL is larger than WHH, a behavior similar to that reported for SC CaFe$_{1.94}$Co$_{0.06}$As$_2$.  Furthermore, the high temperature tail (red dashed lines) in our $H-T$ curve is typical for $H_{c2}(T)$ in the iron-based superconductors.\cite{Park:2008ff} The fact that  $T_{\rm peak}$ is suppressed with increasing current density further suggests that the superconducting filaments are weakly pinned.

It is worth noting that this low temperature anomaly around 10 K is about the same for the maximum $T_c$ of the material with applied pressure.\cite{AmesCa122pressure}   Furthermore, it is present not only in CaFe$_2$As$_2$, but also in the other members of the AFe$_2$As$_2$ family. For example, BaFe$_2$As$_2$ at ambient pressure displays a broad maximum in the resistivity around 20 K, and pressure studies show that even very moderate uniaxial stress can induce at least filamentary SC and a maximum $T_c$ around that temperature.\cite{Duncan:2010hl}

\subsection{Summary}
Enhanced superfluid density at DWs may be a natural consequence of the coupling between SC and AFM orders. Indeed, model calculations exhibit an enhanced local density of states and superconducting order at twin boundaries in the iron pnictides.\cite{DWsFePnictides2011theory}  It is therefore not surprising that superconductivity could nucleate at \emph{antiphase} domain boundaries. However, the fact that the emergence of SC coincides with the freezing of the AFM DWs is striking, and implies that the former is driven by the latter.  In other words, defects in the AFM background are pinned by the emergence of SC order, which would be unstable if the DWs were mobile.
\\
\\
\textbf{Acknowledgments}
We thank S. Roeske at the UCD Electron Microprobe lab and  T. Devereaux, P. Hirschfeld, L. Kemper, K. Kovnir and R. Singh for fruitful discussions. We acknowledge financial support by MEXT-CT-2006-039047, EURYI, National Research Foundation, Singapore and the National Science Foundation under Grant No. DMR-1005393 and DMR-1006606. TH acknowledges support from ICAM Branches Cost Sharing Fund of the Institute for Complex Adaptive Matter and NSF Grant DMR-0844115.
\\
\section{Appendix}

\subsection{NMR Spectra of Domain Walls}

In the presence of either a twin boundary DW or an antiphase boundary DW, the local magnetic order of the Fe moments will be perturbed from the equilibrium antiferromagnetic structure.  This perturbations will modify the local hyperfine field at the As sites, which is given by the vector sum of contributions from the four nearest neighbor Fe sites.  The hyperfine coupling is given by:
\begin{equation}
\mathcal{H}_{\rm hf} = \gamma\hbar\mathbf{\hat{I}}\cdot\sum_{i\in nn}\mathbb{B}_i\cdot\mathbf{S}(\mathbf{r}_i),
\end{equation}
where the sum is over the four nearest neighbor Fe spins $\mathbf{S}(\mathbf{r}_i)$, and the components of the hyperfine tensor   $\mathbb{B}$ are given for CaFe$_2$As$_2$ by: $B_{aa} = B_{bb} = 5.8$ kOe/$\mu_B$, $B_{cc} = 6.0$ kOe/$\mu_B$  and $B_{ac} = 8.2$ kOe/$\mu_B$.\cite{takigawa2008,DioguardiBa122}

\begin{figure}
\centering
\includegraphics[trim=0cm 0cm 0cm 0cm, clip=true, width=0.5\textwidth]{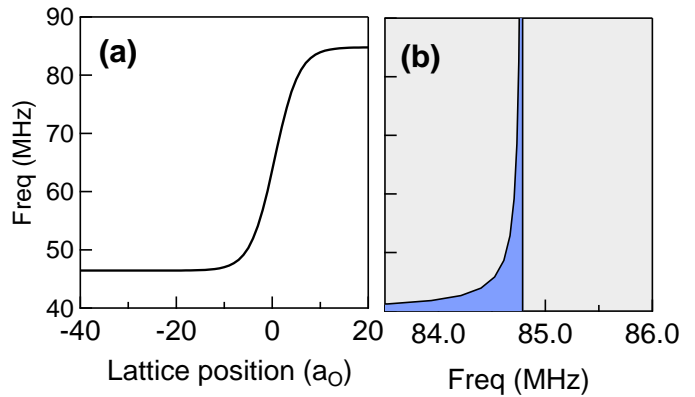}
\caption{ \label{fig:supmat} (Color online) (a) The resonance frequency of the upper satellite of the As along the $b$ direction upon crossing an antiphase DW oriented along the $a$ direction, assuming a DW thickness $\delta = 5a$. (b) The expected lineshape of the As resonance for such a DW.}
\end{figure}

\begin{figure}
\centering
\includegraphics[trim=0cm 0cm 0cm 0cm, clip=true, width=0.5\textwidth]{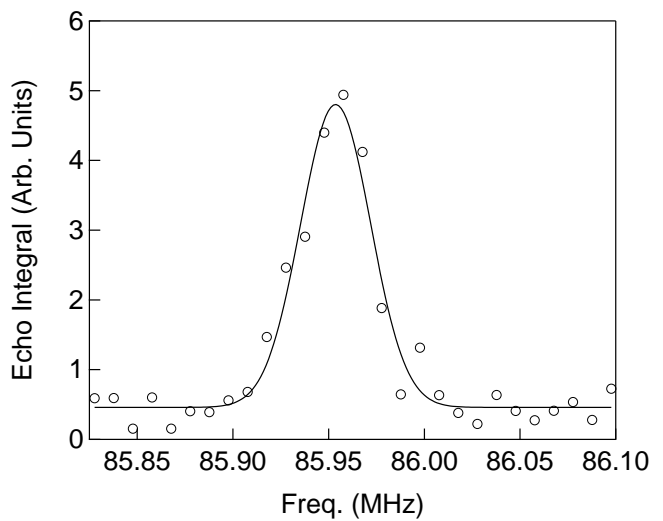}
\caption{ \label{fig:spec} The central resonance of the As in a field of 11.72 T at 300K.  The solid line is a fit to a Guassian with second moment $\sigma = 18(1)$ kHz.}
\end{figure}

The exact structure of the ordered Fe moments in the vicinity of the domain walls depends on details of the magnetic model and the exchange couplings, which has not been completely resolved.  However, in general one can expect the magnetic order to recover to the equilibrium structure within several lattice constants.  As a result, the As hyperfine field will have a similar recovery length at which point the hyperfine field will point alternately along $\pm\hat{c}$.  Within the distance $\delta$ from the boundary, $\mathbf{H}_{\rm hyp}$ will be tilted away from the $\hat{c}$-axis and have components in the $ab$-plane.  This arrangement will modify the resonance frequency, which is proportional to the vector sum $|\mathbf{H}_0 + \mathbf{H}_{\rm hyp}|$.  We therefore approximate the spatial dependence of the resonance frequency along a direction perpendicular to the DW as $\omega(x) = \gamma H_0 + \omega_0\tanh(x/\delta)$ where $\delta$ is the DW width,  $\omega_0 =\gamma H_{\rm hyp} = 19.2$ MHz, and $\gamma H_0 = 65.4$ MHz.   For example, this would be the case for an antiphase DW oriented along the $ab$ direction, where $x$ points along the $b$-direction (see Fig. 1(e)).   This dependence is shown in Fig. \ref{fig:supmat}(a). For such a one-dimensional model, the histogram of resonance frequencies is shown in Fig. \ref{fig:supmat}(b).

\begin{figure}
\centering
\includegraphics[trim=0cm 0cm 0cm 0cm, clip=true, width=0.5\textwidth]{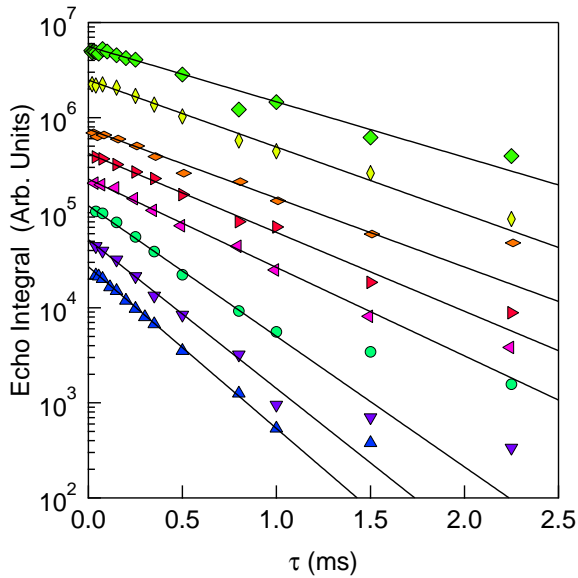}
\caption{ \label{fig:edk} (Color online) Echo decay curves for the As central transition in a field of 5 T for the external field aligned perpendicular to the $c$ axis. The echo integral is shown as a function of pulse spacing, $\tau$, between the 90$^{\circ}$ and 180$^{\circ}$ pulses at several temperatures, offset vertically for clarity.  The temperatures are (in order of increasing vertical offset): 4.5 K, 6 K, 8 K, 12.5 K, 15 K, 17.5 K, 20 K, 25 K, and 30 K. The solid lines are fits to as described in the text.}
\end{figure}


In order to model the spectrum, we first calculate the histogram:

\begin{equation}
P(\omega) = \frac{1}{\Lambda} \int_{-\Lambda/2}^{+\Lambda/2}\delta\left(\omega - \omega(x)\right)dx,
\end{equation}
where $\Lambda$ is the domain size.  This distribution, however, should be convoluted with a Gaussian in order to reproduce the experimental data.  The width of this Gaussian is determined in part by the intrinsic width of the spectrum, as well as the disorder of the domain structure.  In this case, the spectrum is given by:
\begin{eqnarray}
S(\omega) &=& \int_{-\infty}^{+\infty} P(\omega')e^{-(\omega - \omega')^2/2\sigma^2}d\omega' \\
&=& \frac{\delta}{\Lambda}\int_{-\Lambda/2\delta}^{+\Lambda/2\delta}\exp\left[-\frac{(\omega - \omega_0\tanh y)^2}{2\sigma^2}\right]dy.
\end{eqnarray}
We fit the experimental spectra to this function with $\omega_0$, $\sigma$ and $\delta/\Omega$ as variable parameters.  We find that $\sigma\approx 62$ kHz, about 3.5 times the paramagnetic state width.

The relative weight contained in the low frequency tail from the domain walls is given roughly by $f= 2\delta/\Lambda$. For the case of twin domains, transmission electron microscopy and optical measurements find domain sizes on the order of 40 - 400 nm.\cite{prozorovDW122}  Our observation of $f\approx 0.2$ implies that $\delta$ is on the order of 8 - 80 nm, or $10a_{\rm O}$ - $100a_{\rm O}$, where $a_{\rm O} = 0.554$ nm is the orthorhombic unit cell length. It is important to note that at a twin boundary, the structural distortion can be fairly small, but the magnetic response can be extended spatially depending on the details of the magnetic couplings.

\subsection{Lineshape in Paramagnetic State}

Fig. \ref{fig:spec} shows the spectrum of the upper satellite above $T_N$, which shows a Gaussian form with second moment 18(1) kHz.

\subsection{Spin Echo Decay Curves}

The echo decay is shown in Fig. \ref{fig:edk} for a series of temperatures, as well as fits to the function $M(2\tau) = M_0\exp(-2\tau/T_2)$.   The echo decay clearly shows exponential behavior, and the rate $T_{2}^{-1}$  increases with decreasing temperature. The temperature dependence is shown in Fig. 1(d).



$^{\star}$Permanent address: Institute of Physics, Chinese Academy of Sciences, Beijing 100190, China. Email: hxiao@iphy.ac.cn.
$^{\dagger}$Permanent address: Institute of Microsystem and Information Technology, Chinese Academy of Sciences, Shanghhai, China.

\begin{thebibliography}{36}%
\makeatletter
\providecommand \@ifxundefined [1]{%
 \@ifx{#1\undefined}
}%
\providecommand \@ifnum [1]{%
 \ifnum #1\expandafter \@firstoftwo
 \else \expandafter \@secondoftwo
 \fi
}%
\providecommand \@ifx [1]{%
 \ifx #1\expandafter \@firstoftwo
 \else \expandafter \@secondoftwo
 \fi
}%
\providecommand \natexlab [1]{#1}%
\providecommand \enquote  [1]{``#1''}%
\providecommand \bibnamefont  [1]{#1}%
\providecommand \bibfnamefont [1]{#1}%
\providecommand \citenamefont [1]{#1}%
\providecommand \href@noop [0]{\@secondoftwo}%
\providecommand \href [0]{\begingroup \@sanitize@url \@href}%
\providecommand \@href[1]{\@@startlink{#1}\@@href}%
\providecommand \@@href[1]{\endgroup#1\@@endlink}%
\providecommand \@sanitize@url [0]{\catcode `\\12\catcode `\$12\catcode
  `\&12\catcode `\#12\catcode `\^12\catcode `\_12\catcode `\%12\relax}%
\providecommand \@@startlink[1]{}%
\providecommand \@@endlink[0]{}%
\providecommand \url  [0]{\begingroup\@sanitize@url \@url }%
\providecommand \@url [1]{\endgroup\@href {#1}{\urlprefix }}%
\providecommand \urlprefix  [0]{URL }%
\providecommand \Eprint [0]{\href }%
\providecommand \doibase [0]{http://dx.doi.org/}%
\providecommand \selectlanguage [0]{\@gobble}%
\providecommand \bibinfo  [0]{\@secondoftwo}%
\providecommand \bibfield  [0]{\@secondoftwo}%
\providecommand \translation [1]{[#1]}%
\providecommand \BibitemOpen [0]{}%
\providecommand \bibitemStop [0]{}%
\providecommand \bibitemNoStop [0]{.\EOS\space}%
\providecommand \EOS [0]{\spacefactor3000\relax}%
\providecommand \BibitemShut  [1]{\csname bibitem#1\endcsname}%
\let\auto@bib@innerbib\@empty
\bibitem [{\citenamefont {Kamihara}\ \emph {et~al.}(2008)\citenamefont
  {Kamihara}, \citenamefont {Watanabe}, \citenamefont {Hirano},\ and\
  \citenamefont {Hosono}}]{LaOFFeAsJACS}%
  \BibitemOpen
  \bibfield  {author} {\bibinfo {author} {\bibfnamefont {Y.}~\bibnamefont
  {Kamihara}}, \bibinfo {author} {\bibfnamefont {T.}~\bibnamefont {Watanabe}},
  \bibinfo {author} {\bibfnamefont {M.}~\bibnamefont {Hirano}}, \ and\ \bibinfo
  {author} {\bibfnamefont {H.}~\bibnamefont {Hosono}},\ }\href {\doibase
  10.1038/nature06972} {\bibfield  {journal} {\bibinfo  {journal} {J. Am. Chem.
  Soc.}\ }\textbf {\bibinfo {volume} {130}},\ \bibinfo {pages} {3296} (\bibinfo
  {year} {2008})}\BibitemShut {NoStop}%
\bibitem [{\citenamefont {Canfield}\ and\ \citenamefont
  {Bud'ko}(2010)}]{doping122review}%
  \BibitemOpen
  \bibfield  {author} {\bibinfo {author} {\bibfnamefont {P.~C.}\ \bibnamefont
  {Canfield}}\ and\ \bibinfo {author} {\bibfnamefont {S.~L.}\ \bibnamefont
  {Bud'ko}},\ }\href {\doibase 10.1146/annurev-conmatphys-070909-104041}
  {\bibfield  {journal} {\bibinfo  {journal} {Annu. Rev. Condens. Matter
  Phys.}\ }\textbf {\bibinfo {volume} {1}},\ \bibinfo {pages} {27} (\bibinfo
  {year} {2010})}\BibitemShut {NoStop}%
\bibitem [{\citenamefont {Chubukov}\ \emph {et~al.}(2008)\citenamefont
  {Chubukov}, \citenamefont {Efremov},\ and\ \citenamefont
  {Eremin}}]{EreminChubokovPRL2008}%
  \BibitemOpen
  \bibfield  {author} {\bibinfo {author} {\bibfnamefont {A.~V.}\ \bibnamefont
  {Chubukov}}, \bibinfo {author} {\bibfnamefont {D.~V.}\ \bibnamefont
  {Efremov}}, \ and\ \bibinfo {author} {\bibfnamefont {I.}~\bibnamefont
  {Eremin}},\ }\href {\doibase 10.1103/PhysRevB.78.134512} {\bibfield
  {journal} {\bibinfo  {journal} {Phys. Rev. B}\ }\textbf {\bibinfo {volume}
  {78}},\ \bibinfo {pages} {134512} (\bibinfo {year} {2008})}\BibitemShut
  {NoStop}%
\bibitem [{\citenamefont {Mazin}\ \emph {et~al.}(2008)\citenamefont {Mazin},
  \citenamefont {Singh}, \citenamefont {Johannes},\ and\ \citenamefont
  {Du}}]{MazinPairingPRL2008}%
  \BibitemOpen
  \bibfield  {author} {\bibinfo {author} {\bibfnamefont {I.~I.}\ \bibnamefont
  {Mazin}}, \bibinfo {author} {\bibfnamefont {D.~J.}\ \bibnamefont {Singh}},
  \bibinfo {author} {\bibfnamefont {M.~D.}\ \bibnamefont {Johannes}}, \ and\
  \bibinfo {author} {\bibfnamefont {M.~H.}\ \bibnamefont {Du}},\ }\href
  {\doibase 10.1103/PhysRevLett.101.057003} {\bibfield  {journal} {\bibinfo
  {journal} {Phys. Rev. Lett.}\ }\textbf {\bibinfo {volume} {101}},\ \bibinfo
  {eid} {057003} (\bibinfo {year} {2008})}\BibitemShut {NoStop}%
\bibitem [{\citenamefont {Fernandes}\ and\ \citenamefont
  {Schmalian}(2010)}]{Fernandes2010}%
  \BibitemOpen
  \bibfield  {author} {\bibinfo {author} {\bibfnamefont {R.~M.}\ \bibnamefont
  {Fernandes}}\ and\ \bibinfo {author} {\bibfnamefont {J.}~\bibnamefont
  {Schmalian}},\ }\href {\doibase 10.1103/PhysRevB.82.014521} {\bibfield
  {journal} {\bibinfo  {journal} {Phys. Rev. B}\ }\textbf {\bibinfo {volume}
  {82}},\ \bibinfo {pages} {014521} (\bibinfo {year} {2010})}\BibitemShut
  {NoStop}%
\bibitem [{\citenamefont {Alldredge}\ \emph {et~al.}(2008)\citenamefont
  {Alldredge}, \citenamefont {Lee}, \citenamefont {McElroy}, \citenamefont
  {Wang}, \citenamefont {Fujita}, \citenamefont {Kohsaka}, \citenamefont
  {Taylor}, \citenamefont {Eisaki}, \citenamefont {Uchida}, \citenamefont
  {Hirschfeld},\ and\ \citenamefont {Davis}}]{Alldredge2008}%
  \BibitemOpen
  \bibfield  {author} {\bibinfo {author} {\bibfnamefont {J.~W.}\ \bibnamefont
  {Alldredge}}, \bibinfo {author} {\bibfnamefont {J.}~\bibnamefont {Lee}},
  \bibinfo {author} {\bibfnamefont {K.}~\bibnamefont {McElroy}}, \bibinfo
  {author} {\bibfnamefont {M.}~\bibnamefont {Wang}}, \bibinfo {author}
  {\bibfnamefont {K.}~\bibnamefont {Fujita}}, \bibinfo {author} {\bibfnamefont
  {Y.}~\bibnamefont {Kohsaka}}, \bibinfo {author} {\bibfnamefont
  {C.}~\bibnamefont {Taylor}}, \bibinfo {author} {\bibfnamefont
  {H.}~\bibnamefont {Eisaki}}, \bibinfo {author} {\bibfnamefont
  {S.}~\bibnamefont {Uchida}}, \bibinfo {author} {\bibfnamefont {P.~J.}\
  \bibnamefont {Hirschfeld}}, \ and\ \bibinfo {author} {\bibfnamefont {J.~C.}\
  \bibnamefont {Davis}},\ }\href {http://dx.doi.org/10.1038/nphys917}
  {\bibfield  {journal} {\bibinfo  {journal} {Nat. Phys.}\ }\textbf {\bibinfo
  {volume} {4}},\ \bibinfo {pages} {319} (\bibinfo {year} {2008})}\BibitemShut
  {NoStop}%
\bibitem [{\citenamefont {Pratt}\ \emph {et~al.}(2011)\citenamefont {Pratt},
  \citenamefont {Kim}, \citenamefont {Kreyssig}, \citenamefont {Lee},
  \citenamefont {Tucker}, \citenamefont {Thaler}, \citenamefont {Tian},
  \citenamefont {Zarestky}, \citenamefont {Bud'ko}, \citenamefont {Canfield},
  \citenamefont {Harmon}, \citenamefont {Goldman},\ and\ \citenamefont
  {McQueeney}}]{NSBa122incommensurate}%
  \BibitemOpen
  \bibfield  {author} {\bibinfo {author} {\bibfnamefont {D.~K.}\ \bibnamefont
  {Pratt}}, \bibinfo {author} {\bibfnamefont {M.~G.}\ \bibnamefont {Kim}},
  \bibinfo {author} {\bibfnamefont {A.}~\bibnamefont {Kreyssig}}, \bibinfo
  {author} {\bibfnamefont {Y.~B.}\ \bibnamefont {Lee}}, \bibinfo {author}
  {\bibfnamefont {G.~S.}\ \bibnamefont {Tucker}}, \bibinfo {author}
  {\bibfnamefont {A.}~\bibnamefont {Thaler}}, \bibinfo {author} {\bibfnamefont
  {W.}~\bibnamefont {Tian}}, \bibinfo {author} {\bibfnamefont {J.~L.}\
  \bibnamefont {Zarestky}}, \bibinfo {author} {\bibfnamefont {S.~L.}\
  \bibnamefont {Bud'ko}}, \bibinfo {author} {\bibfnamefont {P.~C.}\
  \bibnamefont {Canfield}}, \bibinfo {author} {\bibfnamefont {B.~N.}\
  \bibnamefont {Harmon}}, \bibinfo {author} {\bibfnamefont {A.~I.}\
  \bibnamefont {Goldman}}, \ and\ \bibinfo {author} {\bibfnamefont {R.~J.}\
  \bibnamefont {McQueeney}},\ }\href {\doibase 10.1103/PhysRevLett.106.257001}
  {\bibfield  {journal} {\bibinfo  {journal} {Phys. Rev. Lett.}\ }\textbf
  {\bibinfo {volume} {106}},\ \bibinfo {pages} {257001} (\bibinfo {year}
  {2011})}\BibitemShut {NoStop}%
\bibitem [{\citenamefont {Zhang}\ \emph {et~al.}(2002)\citenamefont {Zhang},
  \citenamefont {Demler},\ and\ \citenamefont {Sachdev}}]{demlerSCandSDW}%
  \BibitemOpen
  \bibfield  {author} {\bibinfo {author} {\bibfnamefont {Y.}~\bibnamefont
  {Zhang}}, \bibinfo {author} {\bibfnamefont {E.}~\bibnamefont {Demler}}, \
  and\ \bibinfo {author} {\bibfnamefont {S.}~\bibnamefont {Sachdev}},\ }\href
  {\doibase 10.1103/PhysRevB.66.094501} {\bibfield  {journal} {\bibinfo
  {journal} {Phys. Rev. B}\ }\textbf {\bibinfo {volume} {66}},\ \bibinfo
  {pages} {094501} (\bibinfo {year} {2002})}\BibitemShut {NoStop}%
\bibitem [{\citenamefont {Kitagawa}\ \emph {et~al.}(2009)\citenamefont
  {Kitagawa}, \citenamefont {Katayama}, \citenamefont {Gotou}, \citenamefont
  {Yagi}, \citenamefont {Ohgushi}, \citenamefont {Matsumoto}, \citenamefont
  {Uwatoko},\ and\ \citenamefont {Takigawa}}]{TakigawaSr122pressure}%
  \BibitemOpen
  \bibfield  {author} {\bibinfo {author} {\bibfnamefont {K.}~\bibnamefont
  {Kitagawa}}, \bibinfo {author} {\bibfnamefont {N.}~\bibnamefont {Katayama}},
  \bibinfo {author} {\bibfnamefont {H.}~\bibnamefont {Gotou}}, \bibinfo
  {author} {\bibfnamefont {T.}~\bibnamefont {Yagi}}, \bibinfo {author}
  {\bibfnamefont {K.}~\bibnamefont {Ohgushi}}, \bibinfo {author} {\bibfnamefont
  {T.}~\bibnamefont {Matsumoto}}, \bibinfo {author} {\bibfnamefont
  {Y.}~\bibnamefont {Uwatoko}}, \ and\ \bibinfo {author} {\bibfnamefont
  {M.}~\bibnamefont {Takigawa}},\ }\href {\doibase
  10.1103/PhysRevLett.103.257002} {\bibfield  {journal} {\bibinfo  {journal}
  {Phys. Rev. Lett.}\ }\textbf {\bibinfo {volume} {103}},\ \bibinfo {pages}
  {257002} (\bibinfo {year} {2009})}\BibitemShut {NoStop}%
\bibitem [{\citenamefont {Kalisky}\ \emph {et~al.}(2010)\citenamefont
  {Kalisky}, \citenamefont {Kirtley}, \citenamefont {Analytis}, \citenamefont
  {Chu}, \citenamefont {Vailionis}, \citenamefont {Fisher},\ and\ \citenamefont
  {Moler}}]{MolerStripesBaFe2As2PRB}%
  \BibitemOpen
  \bibfield  {author} {\bibinfo {author} {\bibfnamefont {B.}~\bibnamefont
  {Kalisky}}, \bibinfo {author} {\bibfnamefont {J.~R.}\ \bibnamefont
  {Kirtley}}, \bibinfo {author} {\bibfnamefont {J.~G.}\ \bibnamefont
  {Analytis}}, \bibinfo {author} {\bibfnamefont {J.-H.}\ \bibnamefont {Chu}},
  \bibinfo {author} {\bibfnamefont {A.}~\bibnamefont {Vailionis}}, \bibinfo
  {author} {\bibfnamefont {I.~R.}\ \bibnamefont {Fisher}}, \ and\ \bibinfo
  {author} {\bibfnamefont {K.~A.}\ \bibnamefont {Moler}},\ }\href {\doibase
  10.1103/PhysRevB.81.184513} {\bibfield  {journal} {\bibinfo  {journal} {Phys.
  Rev. B}\ }\textbf {\bibinfo {volume} {81}},\ \bibinfo {pages} {184513}
  (\bibinfo {year} {2010})}\BibitemShut {NoStop}%
\bibitem [{\citenamefont {Rotter}\ \emph {et~al.}(2008)\citenamefont {Rotter},
  \citenamefont {Tegel},\ and\ \citenamefont {Johrendt}}]{johrendtPRL2008}%
  \BibitemOpen
  \bibfield  {author} {\bibinfo {author} {\bibfnamefont {M.}~\bibnamefont
  {Rotter}}, \bibinfo {author} {\bibfnamefont {M.}~\bibnamefont {Tegel}}, \
  and\ \bibinfo {author} {\bibfnamefont {D.}~\bibnamefont {Johrendt}},\ }\href
  {\doibase 10.1103/PhysRevLett.101.107006} {\bibfield  {journal} {\bibinfo
  {journal} {Phys. Rev. Lett.}\ }\textbf {\bibinfo {volume} {101}},\ \bibinfo
  {pages} {107006} (\bibinfo {year} {2008})}\BibitemShut {NoStop}%
\bibitem [{\citenamefont {Goldman}\ \emph {et~al.}(2009)\citenamefont
  {Goldman}, \citenamefont {Kreyssig}, \citenamefont {Proke\v{s}},
  \citenamefont {Pratt}, \citenamefont {Argyriou}, \citenamefont {Lynn},
  \citenamefont {Nandi}, \citenamefont {Kimber}, \citenamefont {Chen},
  \citenamefont {Lee}, \citenamefont {Samolyuk}, \citenamefont {{a}o},
  \citenamefont {Poulton}, \citenamefont {Bud'ko}, \citenamefont {Ni},
  \citenamefont {Canfield}, \citenamefont {Harmon},\ and\ \citenamefont
  {McQueeney}}]{Goldman2008PRB}%
  \BibitemOpen
  \bibfield  {author} {\bibinfo {author} {\bibfnamefont {A.~I.}\ \bibnamefont
  {Goldman}}, \bibinfo {author} {\bibfnamefont {A.}~\bibnamefont {Kreyssig}},
  \bibinfo {author} {\bibfnamefont {K.}~\bibnamefont {Proke\v{s}}}, \bibinfo
  {author} {\bibfnamefont {D.~K.}\ \bibnamefont {Pratt}}, \bibinfo {author}
  {\bibfnamefont {D.~N.}\ \bibnamefont {Argyriou}}, \bibinfo {author}
  {\bibfnamefont {J.~W.}\ \bibnamefont {Lynn}}, \bibinfo {author}
  {\bibfnamefont {S.}~\bibnamefont {Nandi}}, \bibinfo {author} {\bibfnamefont
  {S.~A.~J.}\ \bibnamefont {Kimber}}, \bibinfo {author} {\bibfnamefont
  {Y.}~\bibnamefont {Chen}}, \bibinfo {author} {\bibfnamefont {Y.~B.}\
  \bibnamefont {Lee}}, \bibinfo {author} {\bibfnamefont {G.}~\bibnamefont
  {Samolyuk}}, \bibinfo {author} {\bibfnamefont {J.~B.~L.}\ \bibnamefont
  {{a}o}}, \bibinfo {author} {\bibfnamefont {S.~J.}\ \bibnamefont {Poulton}},
  \bibinfo {author} {\bibfnamefont {S.~L.}\ \bibnamefont {Bud'ko}}, \bibinfo
  {author} {\bibfnamefont {N.}~\bibnamefont {Ni}}, \bibinfo {author}
  {\bibfnamefont {P.~C.}\ \bibnamefont {Canfield}}, \bibinfo {author}
  {\bibfnamefont {B.~N.}\ \bibnamefont {Harmon}}, \ and\ \bibinfo {author}
  {\bibfnamefont {R.~J.}\ \bibnamefont {McQueeney}},\ }\href {\doibase
  10.1103/PhysRevB.79.024513} {\bibfield  {journal} {\bibinfo  {journal} {Phys.
  Rev. B}\ }\textbf {\bibinfo {volume} {79}},\ \bibinfo {eid} {024513}
  (\bibinfo {year} {2009})}\BibitemShut {NoStop}%
\bibitem [{\citenamefont {Monthoux}\ \emph {et~al.}(2007)\citenamefont
  {Monthoux}, \citenamefont {Pines},\ and\ \citenamefont
  {Lonzarich}}]{MonthouxPinesReview}%
  \BibitemOpen
  \bibfield  {author} {\bibinfo {author} {\bibfnamefont {P.}~\bibnamefont
  {Monthoux}}, \bibinfo {author} {\bibfnamefont {D.}~\bibnamefont {Pines}}, \
  and\ \bibinfo {author} {\bibfnamefont {G.~G.}\ \bibnamefont {Lonzarich}},\
  }\href {\doibase 10.1038/nature06480} {\bibfield  {journal} {\bibinfo
  {journal} {Nature}\ }\textbf {\bibinfo {volume} {450}},\ \bibinfo {pages}
  {1177} (\bibinfo {year} {2007})}\BibitemShut {NoStop}%
\bibitem [{\citenamefont {Baek}\ \emph {et~al.}(2009)\citenamefont {Baek},
  \citenamefont {Lee}, \citenamefont {Brown}, \citenamefont {Curro},
  \citenamefont {Bauer}, \citenamefont {Ronning}, \citenamefont {Park},\ and\
  \citenamefont {Thompson}}]{seunghoCa122PRL}%
  \BibitemOpen
  \bibfield  {author} {\bibinfo {author} {\bibfnamefont {S.-H.}\ \bibnamefont
  {Baek}}, \bibinfo {author} {\bibfnamefont {H.}~\bibnamefont {Lee}}, \bibinfo
  {author} {\bibfnamefont {S.~E.}\ \bibnamefont {Brown}}, \bibinfo {author}
  {\bibfnamefont {N.~J.}\ \bibnamefont {Curro}}, \bibinfo {author}
  {\bibfnamefont {E.~D.}\ \bibnamefont {Bauer}}, \bibinfo {author}
  {\bibfnamefont {F.}~\bibnamefont {Ronning}}, \bibinfo {author} {\bibfnamefont
  {T.}~\bibnamefont {Park}}, \ and\ \bibinfo {author} {\bibfnamefont {J.~D.}\
  \bibnamefont {Thompson}},\ }\href {\doibase 10.1103/PhysRevLett.102.227601}
  {\bibfield  {journal} {\bibinfo  {journal} {Phys. Rev. Lett.}\ }\textbf
  {\bibinfo {volume} {102}},\ \bibinfo {eid} {227601} (\bibinfo {year}
  {2009})}\BibitemShut {NoStop}%
\bibitem [{\citenamefont {Torikachvili}\ \emph
  {et~al.}(2008{\natexlab{a}})\citenamefont {Torikachvili}, \citenamefont
  {Bud'ko}, \citenamefont {Ni},\ and\ \citenamefont
  {Canfield}}]{AmesCa122pressure}%
  \BibitemOpen
  \bibfield  {author} {\bibinfo {author} {\bibfnamefont {M.~S.}\ \bibnamefont
  {Torikachvili}}, \bibinfo {author} {\bibfnamefont {S.~L.}\ \bibnamefont
  {Bud'ko}}, \bibinfo {author} {\bibfnamefont {N.}~\bibnamefont {Ni}}, \ and\
  \bibinfo {author} {\bibfnamefont {P.~C.}\ \bibnamefont {Canfield}},\ }\href
  {\doibase 10.1103/PhysRevLett.101.057006} {\bibfield  {journal} {\bibinfo
  {journal} {Phys. Rev. Lett.}\ }\textbf {\bibinfo {volume} {101}},\ \bibinfo
  {pages} {057006} (\bibinfo {year} {2008}{\natexlab{a}})}\BibitemShut
  {NoStop}%
\bibitem [{\citenamefont {Alireza}\ \emph {et~al.}(2009)\citenamefont
  {Alireza}, \citenamefont {Ko}, \citenamefont {Gillett}, \citenamefont
  {Petrone}, \citenamefont {Cole}, \citenamefont {Lonzarich},\ and\
  \citenamefont {Sebastian}}]{Cambridge122pressure2009}%
  \BibitemOpen
  \bibfield  {author} {\bibinfo {author} {\bibfnamefont {P.~L.}\ \bibnamefont
  {Alireza}}, \bibinfo {author} {\bibfnamefont {Y.~T.~C.}\ \bibnamefont {Ko}},
  \bibinfo {author} {\bibfnamefont {J.}~\bibnamefont {Gillett}}, \bibinfo
  {author} {\bibfnamefont {C.~M.}\ \bibnamefont {Petrone}}, \bibinfo {author}
  {\bibfnamefont {J.~M.}\ \bibnamefont {Cole}}, \bibinfo {author}
  {\bibfnamefont {G.~G.}\ \bibnamefont {Lonzarich}}, \ and\ \bibinfo {author}
  {\bibfnamefont {S.~E.}\ \bibnamefont {Sebastian}},\ }\href
  {http://stacks.iop.org/0953-8984/21/i=1/a=012208} {\bibfield  {journal}
  {\bibinfo  {journal} {J. Phys.: Condens. Matter}\ }\textbf {\bibinfo {volume}
  {21}},\ \bibinfo {pages} {012208} (\bibinfo {year} {2009})}\BibitemShut
  {NoStop}%
\bibitem [{\citenamefont {Park}\ and\ \citenamefont
  {Thompson}(2009)}]{ParkThompsonRh115review}%
  \BibitemOpen
  \bibfield  {author} {\bibinfo {author} {\bibfnamefont {T.}~\bibnamefont
  {Park}}\ and\ \bibinfo {author} {\bibfnamefont {J.~D.}\ \bibnamefont
  {Thompson}},\ }\href {\doibase 10.1088/1367-2630/11/5/055062} {\bibfield
  {journal} {\bibinfo  {journal} {New J. Phys.}\ }\textbf {\bibinfo {volume}
  {11}},\ \bibinfo {pages} {055062} (\bibinfo {year} {2009})}\BibitemShut
  {NoStop}%
\bibitem [{\citenamefont {Saha}\ \emph {et~al.}(2009)\citenamefont {Saha},
  \citenamefont {Butch}, \citenamefont {Kirshenbaum}, \citenamefont
  {Paglione},\ and\ \citenamefont {Zavalij}}]{paglione2011}%
  \BibitemOpen
  \bibfield  {author} {\bibinfo {author} {\bibfnamefont {S.~R.}\ \bibnamefont
  {Saha}}, \bibinfo {author} {\bibfnamefont {N.~P.}\ \bibnamefont {Butch}},
  \bibinfo {author} {\bibfnamefont {K.}~\bibnamefont {Kirshenbaum}}, \bibinfo
  {author} {\bibfnamefont {J.}~\bibnamefont {Paglione}}, \ and\ \bibinfo
  {author} {\bibfnamefont {P.~Y.}\ \bibnamefont {Zavalij}},\ }\href {\doibase
  10.1103/PhysRevLett.103.037005} {\bibfield  {journal} {\bibinfo  {journal}
  {Phys. Rev. Lett.}\ }\textbf {\bibinfo {volume} {103}},\ \bibinfo {pages}
  {037005} (\bibinfo {year} {2009})}\BibitemShut {NoStop}%
\bibitem [{\citenamefont {Curro}\ \emph {et~al.}(2009)\citenamefont {Curro},
  \citenamefont {Dioguardi}, \citenamefont {ApRoberts-Warren}, \citenamefont
  {Shockley},\ and\ \citenamefont {Klavins}}]{adamCaFe2As2}%
  \BibitemOpen
  \bibfield  {author} {\bibinfo {author} {\bibfnamefont {N.~J.}\ \bibnamefont
  {Curro}}, \bibinfo {author} {\bibfnamefont {A.~P.}\ \bibnamefont
  {Dioguardi}}, \bibinfo {author} {\bibfnamefont {N.}~\bibnamefont
  {ApRoberts-Warren}}, \bibinfo {author} {\bibfnamefont {A.~C.}\ \bibnamefont
  {Shockley}}, \ and\ \bibinfo {author} {\bibfnamefont {P.}~\bibnamefont
  {Klavins}},\ }\href {http://stacks.iop.org/1367-2630/11/075004} {\bibfield
  {journal} {\bibinfo  {journal} {New J. Phys.}\ }\textbf {\bibinfo {volume}
  {11}},\ \bibinfo {pages} {075004 (10pp)} (\bibinfo {year}
  {2009})}\BibitemShut {NoStop}%
\bibitem [{\citenamefont {Mazin}\ and\ \citenamefont
  {Johannes}(2009)}]{MazinNature2009}%
  \BibitemOpen
  \bibfield  {author} {\bibinfo {author} {\bibfnamefont {I.~I.}\ \bibnamefont
  {Mazin}}\ and\ \bibinfo {author} {\bibfnamefont {M.~D.}\ \bibnamefont
  {Johannes}},\ }\href {http://dx.doi.org/10.1038/nphys1160} {\bibfield
  {journal} {\bibinfo  {journal} {Nat. Phys.}\ }\textbf {\bibinfo {volume}
  {5}},\ \bibinfo {pages} {141} (\bibinfo {year} {2009})}\BibitemShut {NoStop}%
\bibitem [{\citenamefont {Ahn}\ \emph {et~al.}(2006)\citenamefont {Ahn},
  \citenamefont {Bhattacharya}, \citenamefont {Ventra}, \citenamefont
  {Eckstein}, \citenamefont {Frisbie}, \citenamefont {Gershenson},
  \citenamefont {Goldman}, \citenamefont {Inoue}, \citenamefont {Mannhart},
  \citenamefont {Millis}, \citenamefont {Morpurgo}, \citenamefont {Natelson},\
  and\ \citenamefont {Triscone}}]{MillisReview}%
  \BibitemOpen
  \bibfield  {author} {\bibinfo {author} {\bibfnamefont {C.~H.}\ \bibnamefont
  {Ahn}}, \bibinfo {author} {\bibfnamefont {A.}~\bibnamefont {Bhattacharya}},
  \bibinfo {author} {\bibfnamefont {M.~D.}\ \bibnamefont {Ventra}}, \bibinfo
  {author} {\bibfnamefont {J.~N.}\ \bibnamefont {Eckstein}}, \bibinfo {author}
  {\bibfnamefont {C.~D.}\ \bibnamefont {Frisbie}}, \bibinfo {author}
  {\bibfnamefont {M.~E.}\ \bibnamefont {Gershenson}}, \bibinfo {author}
  {\bibfnamefont {A.~M.}\ \bibnamefont {Goldman}}, \bibinfo {author}
  {\bibfnamefont {I.~H.}\ \bibnamefont {Inoue}}, \bibinfo {author}
  {\bibfnamefont {J.}~\bibnamefont {Mannhart}}, \bibinfo {author}
  {\bibfnamefont {A.~J.}\ \bibnamefont {Millis}}, \bibinfo {author}
  {\bibfnamefont {A.~F.}\ \bibnamefont {Morpurgo}}, \bibinfo {author}
  {\bibfnamefont {D.}~\bibnamefont {Natelson}}, \ and\ \bibinfo {author}
  {\bibfnamefont {J.-M.}\ \bibnamefont {Triscone}},\ }\href {\doibase
  10.1103/RevModPhys.78.1185} {\bibfield  {journal} {\bibinfo  {journal} {Rev.
  Mod. Phys.}\ }\textbf {\bibinfo {volume} {78}},\ \bibinfo {eid} {1185}
  (\bibinfo {year} {2006})}\BibitemShut {NoStop}%
\bibitem [{\citenamefont {Jiang}\ \emph {et~al.}(1997)\citenamefont {Jiang},
  \citenamefont {Baldwin}, \citenamefont {Levin}, \citenamefont {Stein},
  \citenamefont {Almasan}, \citenamefont {Gajewski}, \citenamefont {Han},\ and\
  \citenamefont {Maple}}]{Jiang:1997ee}%
  \BibitemOpen
  \bibfield  {author} {\bibinfo {author} {\bibfnamefont {C.~N.}\ \bibnamefont
  {Jiang}}, \bibinfo {author} {\bibfnamefont {A.~R.}\ \bibnamefont {Baldwin}},
  \bibinfo {author} {\bibfnamefont {G.~A.}\ \bibnamefont {Levin}}, \bibinfo
  {author} {\bibfnamefont {T.}~\bibnamefont {Stein}}, \bibinfo {author}
  {\bibfnamefont {C.~C.}\ \bibnamefont {Almasan}}, \bibinfo {author}
  {\bibfnamefont {D.~A.}\ \bibnamefont {Gajewski}}, \bibinfo {author}
  {\bibfnamefont {S.~H.}\ \bibnamefont {Han}}, \ and\ \bibinfo {author}
  {\bibfnamefont {M.~B.}\ \bibnamefont {Maple}},\ }\href {\doibase
  10.1103/PhysRevB.55.R3390} {\bibfield  {journal} {\bibinfo  {journal} {Phys.
  Rev. B}\ }\textbf {\bibinfo {volume} {55}},\ \bibinfo {pages} {R3390}
  (\bibinfo {year} {1997})}\BibitemShut {NoStop}%
\bibitem [{\citenamefont {Abragam}(1961)}]{abragambook}%
  \BibitemOpen
  \bibfield  {author} {\bibinfo {author} {\bibfnamefont {A.}~\bibnamefont
  {Abragam}},\ }\href@noop {} {\emph {\bibinfo {title} {The Principles of
  Nuclear Magnetism}}}\ (\bibinfo  {publisher} {Oxford University Press,
  Oxford},\ \bibinfo {year} {1961})\BibitemShut {NoStop}%
\bibitem [{\citenamefont {Pennington}\ and\ \citenamefont
  {Slichter}(1991)}]{PenningtonSlichterT2G}%
  \BibitemOpen
  \bibfield  {author} {\bibinfo {author} {\bibfnamefont {C.~H.}\ \bibnamefont
  {Pennington}}\ and\ \bibinfo {author} {\bibfnamefont {C.~P.}\ \bibnamefont
  {Slichter}},\ }\href {\doibase 10.1103/PhysRevLett.66.381} {\bibfield
  {journal} {\bibinfo  {journal} {Phys. Rev. Lett.}\ }\textbf {\bibinfo
  {volume} {66}},\ \bibinfo {pages} {381} (\bibinfo {year} {1991})}\BibitemShut
  {NoStop}%
\bibitem [{\citenamefont {Tanatar}\ \emph {et~al.}(2009)\citenamefont
  {Tanatar}, \citenamefont {Kreyssig}, \citenamefont {Nandi}, \citenamefont
  {Ni}, \citenamefont {Bud'ko}, \citenamefont {Canfield}, \citenamefont
  {Goldman},\ and\ \citenamefont {Prozorov}}]{prozorovDW122}%
  \BibitemOpen
  \bibfield  {author} {\bibinfo {author} {\bibfnamefont {M.~A.}\ \bibnamefont
  {Tanatar}}, \bibinfo {author} {\bibfnamefont {A.}~\bibnamefont {Kreyssig}},
  \bibinfo {author} {\bibfnamefont {S.}~\bibnamefont {Nandi}}, \bibinfo
  {author} {\bibfnamefont {N.}~\bibnamefont {Ni}}, \bibinfo {author}
  {\bibfnamefont {S.~L.}\ \bibnamefont {Bud'ko}}, \bibinfo {author}
  {\bibfnamefont {P.~C.}\ \bibnamefont {Canfield}}, \bibinfo {author}
  {\bibfnamefont {A.~I.}\ \bibnamefont {Goldman}}, \ and\ \bibinfo {author}
  {\bibfnamefont {R.}~\bibnamefont {Prozorov}},\ }\href {\doibase
  10.1103/PhysRevB.79.180508} {\bibfield  {journal} {\bibinfo  {journal} {Phys.
  Rev. B}\ }\textbf {\bibinfo {volume} {79}},\ \bibinfo {pages} {180508}
  (\bibinfo {year} {2009})}\BibitemShut {NoStop}%
\bibitem [{\citenamefont {Li}\ \emph {et~al.}(2010)\citenamefont {Li},
  \citenamefont {He}, \citenamefont {Li}, \citenamefont {Pan}, \citenamefont
  {Zhang}, \citenamefont {Jin}, \citenamefont {Sefat}, \citenamefont {McGuire},
  \citenamefont {Mandrus}, \citenamefont {Sales},\ and\ \citenamefont
  {Plummer}}]{PlummerBa122STM}%
  \BibitemOpen
  \bibfield  {author} {\bibinfo {author} {\bibfnamefont {G.}~\bibnamefont
  {Li}}, \bibinfo {author} {\bibfnamefont {X.}~\bibnamefont {He}}, \bibinfo
  {author} {\bibfnamefont {A.}~\bibnamefont {Li}}, \bibinfo {author}
  {\bibfnamefont {S.~H.}\ \bibnamefont {Pan}}, \bibinfo {author} {\bibfnamefont
  {J.}~\bibnamefont {Zhang}}, \bibinfo {author} {\bibfnamefont
  {R.}~\bibnamefont {Jin}}, \bibinfo {author} {\bibfnamefont {A.~S.}\
  \bibnamefont {Sefat}}, \bibinfo {author} {\bibfnamefont {M.~A.}\ \bibnamefont
  {McGuire}}, \bibinfo {author} {\bibfnamefont {D.~G.}\ \bibnamefont
  {Mandrus}}, \bibinfo {author} {\bibfnamefont {B.~C.}\ \bibnamefont {Sales}},
  \ and\ \bibinfo {author} {\bibfnamefont {E.~W.}\ \bibnamefont {Plummer}},\
  }\href@noop {}\ \Eprint
  {http://arxiv.org/abs/1006.5907} {arXiv:1006.5907} (\bibinfo {year} {2010}) \BibitemShut {NoStop}%
\bibitem [{\citenamefont {Dioguardi}\ \emph {et~al.}(2010)\citenamefont
  {Dioguardi}, \citenamefont {apRoberts Warren}, \citenamefont {Shockley},
  \citenamefont {Bud'ko}, \citenamefont {Ni}, \citenamefont {Canfield},\ and\
  \citenamefont {Curro}}]{DioguardiBa122}%
  \BibitemOpen
  \bibfield  {author} {\bibinfo {author} {\bibfnamefont {A.~P.}\ \bibnamefont
  {Dioguardi}}, \bibinfo {author} {\bibfnamefont {N.}~\bibnamefont {apRoberts
  Warren}}, \bibinfo {author} {\bibfnamefont {A.~C.}\ \bibnamefont {Shockley}},
  \bibinfo {author} {\bibfnamefont {S.~L.}\ \bibnamefont {Bud'ko}}, \bibinfo
  {author} {\bibfnamefont {N.}~\bibnamefont {Ni}}, \bibinfo {author}
  {\bibfnamefont {P.~C.}\ \bibnamefont {Canfield}}, \ and\ \bibinfo {author}
  {\bibfnamefont {N.~J.}\ \bibnamefont {Curro}},\ }\href {\doibase
  10.1103/PhysRevB.82.140411} {\bibfield  {journal} {\bibinfo  {journal} {Phys.
  Rev. B}\ }\textbf {\bibinfo {volume} {82}},\ \bibinfo {pages} {140411}
  (\bibinfo {year} {2010})}\BibitemShut {NoStop}%
\bibitem [{Note1()}]{Note1}%
  \BibitemOpen
  \bibinfo {note} {See Appendix for details of the
  calculation.}\BibitemShut {Stop}%
\bibitem [{\citenamefont {Ronning}\ \emph {et~al.}(2008)\citenamefont
  {Ronning}, \citenamefont {Klimczuk}, \citenamefont {Bauer}, \citenamefont
  {Volz},\ and\ \citenamefont {Thompson}}]{RonningCaFe2As2discovery}%
  \BibitemOpen
  \bibfield  {author} {\bibinfo {author} {\bibfnamefont {F.}~\bibnamefont
  {Ronning}}, \bibinfo {author} {\bibfnamefont {T.}~\bibnamefont {Klimczuk}},
  \bibinfo {author} {\bibfnamefont {E.~D.}\ \bibnamefont {Bauer}}, \bibinfo
  {author} {\bibfnamefont {H.}~\bibnamefont {Volz}}, \ and\ \bibinfo {author}
  {\bibfnamefont {J.~D.}\ \bibnamefont {Thompson}},\ }\href
  {http://stacks.iop.org/0953-8984/20/322201} {\bibfield  {journal} {\bibinfo
  {journal} {J. Phys.: Condens. Matter}\ }\textbf {\bibinfo {volume} {20}},\
  \bibinfo {pages} {322201 (4pp)} (\bibinfo {year} {2008})}\BibitemShut
  {NoStop}%
\bibitem [{\citenamefont {Ni}\ \emph {et~al.}(2008)\citenamefont {Ni},
  \citenamefont {Nandi}, \citenamefont {Kreyssig}, \citenamefont {Goldman},
  \citenamefont {Mun}, \citenamefont {Bud'ko},\ and\ \citenamefont
  {Canfield}}]{Ni:2008jp}%
  \BibitemOpen
  \bibfield  {author} {\bibinfo {author} {\bibfnamefont {N.}~\bibnamefont
  {Ni}}, \bibinfo {author} {\bibfnamefont {S.}~\bibnamefont {Nandi}}, \bibinfo
  {author} {\bibfnamefont {A.}~\bibnamefont {Kreyssig}}, \bibinfo {author}
  {\bibfnamefont {A.~I.}\ \bibnamefont {Goldman}}, \bibinfo {author}
  {\bibfnamefont {E.~D.}\ \bibnamefont {Mun}}, \bibinfo {author} {\bibfnamefont
  {S.~L.}\ \bibnamefont {Bud'ko}}, \ and\ \bibinfo {author} {\bibfnamefont
  {P.~C.}\ \bibnamefont {Canfield}},\ }\href {\doibase
  10.1103/PhysRevB.78.014523} {\bibfield  {journal} {\bibinfo  {journal} {Phys.
  Rev. B}\ }\textbf {\bibinfo {volume} {78}},\ \bibinfo {pages} {014523}
  (\bibinfo {year} {2008})}\BibitemShut {NoStop}%
\bibitem [{\citenamefont {Torikachvili}\ \emph
  {et~al.}(2008{\natexlab{b}})\citenamefont {Torikachvili}, \citenamefont
  {Bud'ko}, \citenamefont {Ni},\ and\ \citenamefont
  {Canfield}}]{CanfieldPressureCa122}%
  \BibitemOpen
  \bibfield  {author} {\bibinfo {author} {\bibfnamefont {M.~S.}\ \bibnamefont
  {Torikachvili}}, \bibinfo {author} {\bibfnamefont {S.~L.}\ \bibnamefont
  {Bud'ko}}, \bibinfo {author} {\bibfnamefont {N.}~\bibnamefont {Ni}}, \ and\
  \bibinfo {author} {\bibfnamefont {P.~C.}\ \bibnamefont {Canfield}},\ }\href
  {\doibase 10.1103/PhysRevLett.101.057006} {\bibfield  {journal} {\bibinfo
  {journal} {Phys. Rev. Lett.}\ }\textbf {\bibinfo {volume} {101}},\ \bibinfo
  {eid} {057006} (\bibinfo {year} {2008}{\natexlab{b}})}\BibitemShut {NoStop}%
\bibitem [{\citenamefont {Torikachvili}\ \emph {et~al.}(2009)\citenamefont
  {Torikachvili}, \citenamefont {Bud'ko}, \citenamefont {Ni}, \citenamefont
  {Canfield},\ and\ \citenamefont {Hannahs}}]{Torikachvili:2009dp}%
  \BibitemOpen
  \bibfield  {author} {\bibinfo {author} {\bibfnamefont {M.~S.}\ \bibnamefont
  {Torikachvili}}, \bibinfo {author} {\bibfnamefont {S.~L.}\ \bibnamefont
  {Bud'ko}}, \bibinfo {author} {\bibfnamefont {N.}~\bibnamefont {Ni}}, \bibinfo
  {author} {\bibfnamefont {P.~C.}\ \bibnamefont {Canfield}}, \ and\ \bibinfo
  {author} {\bibfnamefont {S.~T.}\ \bibnamefont {Hannahs}},\ }\href {\doibase
  10.1103/PhysRevB.80.014521} {\bibfield  {journal} {\bibinfo  {journal} {Phys.
  Rev. B}\ }\textbf {\bibinfo {volume} {80}},\ \bibinfo {pages} {014521}
  (\bibinfo {year} {2009})}\BibitemShut {NoStop}%
\bibitem [{\citenamefont {Kumar}\ \emph {et~al.}(2009)\citenamefont {Kumar},
  \citenamefont {Nagalakshmi}, \citenamefont {Kulkarni}, \citenamefont
  {Paulose}, \citenamefont {Nigam}, \citenamefont {Dhar},\ and\ \citenamefont
  {Thamizhavel}}]{KumarPRB2009}%
  \BibitemOpen
  \bibfield  {author} {\bibinfo {author} {\bibfnamefont {N.}~\bibnamefont
  {Kumar}}, \bibinfo {author} {\bibfnamefont {R.}~\bibnamefont {Nagalakshmi}},
  \bibinfo {author} {\bibfnamefont {R.}~\bibnamefont {Kulkarni}}, \bibinfo
  {author} {\bibfnamefont {P.~L.}\ \bibnamefont {Paulose}}, \bibinfo {author}
  {\bibfnamefont {A.~K.}\ \bibnamefont {Nigam}}, \bibinfo {author}
  {\bibfnamefont {S.~K.}\ \bibnamefont {Dhar}}, \ and\ \bibinfo {author}
  {\bibfnamefont {A.}~\bibnamefont {Thamizhavel}},\ }\href {\doibase
  10.1103/PhysRevB.79.012504} {\bibfield  {journal} {\bibinfo  {journal} {Phys.
  Rev. B}\ }\textbf {\bibinfo {volume} {79}},\ \bibinfo {pages} {012504}
  (\bibinfo {year} {2009})}\BibitemShut {NoStop}%
\bibitem [{\citenamefont {Park}\ \emph {et~al.}(2008)\citenamefont {Park},
  \citenamefont {Park}, \citenamefont {Lee}, \citenamefont {Klimczuk},
  \citenamefont {Bauer}, \citenamefont {Ronning},\ and\ \citenamefont
  {Thompson}}]{Park:2008ff}%
  \BibitemOpen
  \bibfield  {author} {\bibinfo {author} {\bibfnamefont {T.}~\bibnamefont
  {Park}}, \bibinfo {author} {\bibfnamefont {E.}~\bibnamefont {Park}}, \bibinfo
  {author} {\bibfnamefont {H.}~\bibnamefont {Lee}}, \bibinfo {author}
  {\bibfnamefont {T.}~\bibnamefont {Klimczuk}}, \bibinfo {author}
  {\bibfnamefont {E.~D.}\ \bibnamefont {Bauer}}, \bibinfo {author}
  {\bibfnamefont {F.}~\bibnamefont {Ronning}}, \ and\ \bibinfo {author}
  {\bibfnamefont {J.~D.}\ \bibnamefont {Thompson}},\ }\href@noop {} {\bibfield
  {journal} {\bibinfo  {journal} {J. Phys.: Condens. Matter}\ }\textbf
  {\bibinfo {volume} {20}},\ \bibinfo {pages} {322204} (\bibinfo {year}
  {2008})}\BibitemShut {NoStop}%
\bibitem [{\citenamefont {Duncan}\ \emph {et~al.}(2010)\citenamefont {Duncan},
  \citenamefont {Welzel}, \citenamefont {Harrison}, \citenamefont {Wang},
  \citenamefont {Chen}, \citenamefont {Grosche},\ and\ \citenamefont
  {Niklowitz}}]{Duncan:2010hl}%
  \BibitemOpen
  \bibfield  {author} {\bibinfo {author} {\bibfnamefont {W.~J.}\ \bibnamefont
  {Duncan}}, \bibinfo {author} {\bibfnamefont {O.~P.}\ \bibnamefont {Welzel}},
  \bibinfo {author} {\bibfnamefont {C.}~\bibnamefont {Harrison}}, \bibinfo
  {author} {\bibfnamefont {X.~F.}\ \bibnamefont {Wang}}, \bibinfo {author}
  {\bibfnamefont {X.~H.}\ \bibnamefont {Chen}}, \bibinfo {author}
  {\bibfnamefont {F.~M.}\ \bibnamefont {Grosche}}, \ and\ \bibinfo {author}
  {\bibfnamefont {P.~G.}\ \bibnamefont {Niklowitz}},\ }\href {\doibase
  10.1088/0953-8984/22/5/052201} {\bibfield  {journal} {\bibinfo  {journal} {J.
  Phys.: Condens. Matter}\ }\textbf {\bibinfo {volume} {22}},\ \bibinfo {pages}
  {052201} (\bibinfo {year} {2010})}\BibitemShut {NoStop}%
\bibitem [{\citenamefont {Huang}\ \emph {et~al.}(2011)\citenamefont {Huang},
  \citenamefont {Zhang}, \citenamefont {Zhou},\ and\ \citenamefont
  {Ting}}]{DWsFePnictides2011theory}%
  \BibitemOpen
  \bibfield  {author} {\bibinfo {author} {\bibfnamefont {H.}~\bibnamefont
  {Huang}}, \bibinfo {author} {\bibfnamefont {D.}~\bibnamefont {Zhang}},
  \bibinfo {author} {\bibfnamefont {T.}~\bibnamefont {Zhou}}, \ and\ \bibinfo
  {author} {\bibfnamefont {C.~S.}\ \bibnamefont {Ting}},\ }\href {\doibase
  10.1103/PhysRevB.83.134517} {\bibfield  {journal} {\bibinfo  {journal} {Phys.
  Rev. B}\ }\textbf {\bibinfo {volume} {83}},\ \bibinfo {pages} {134517}
  (\bibinfo {year} {2011})}\BibitemShut {NoStop}%
\bibitem [{\citenamefont {Kitagawa}\ \emph {et~al.}(2008)\citenamefont
  {Kitagawa}, \citenamefont {Katayama}, \citenamefont {Ohgushi}, \citenamefont
  {Yoshida},\ and\ \citenamefont {Takigawa}}]{takigawa2008}%
  \BibitemOpen
  \bibfield  {author} {\bibinfo {author} {\bibfnamefont {K.}~\bibnamefont
  {Kitagawa}}, \bibinfo {author} {\bibfnamefont {N.}~\bibnamefont {Katayama}},
  \bibinfo {author} {\bibfnamefont {K.}~\bibnamefont {Ohgushi}}, \bibinfo
  {author} {\bibfnamefont {M.}~\bibnamefont {Yoshida}}, \ and\ \bibinfo
  {author} {\bibfnamefont {M.}~\bibnamefont {Takigawa}},\ }\href@noop {}
  {\bibfield  {journal} {\bibinfo  {journal} {J. Phys. Soc. Jpn.}\ }\textbf
  {\bibinfo {volume} {77}},\ \bibinfo {pages} {114709} (\bibinfo {year}
  {2008})}\BibitemShut {NoStop}%
\end{thebibliography}
%
\end{document}